\documentclass[journal]{IEEEtran}

\usepackage{amsmath}
\usepackage{amssymb,mathrsfs} 
\usepackage{graphicx}
\usepackage{amsthm}

\usepackage{subfig}

\usepackage{bm}
\usepackage{bbm}
\usepackage{amssymb}  
\usepackage{pifont}
\usepackage{array} 
\usepackage{tabularx} 
\usepackage{epstopdf}
\usepackage{dsfont}
\usepackage{color}
\usepackage{epsfig}
\usepackage{amsthm}
\makeatletter

\newcommand{\Rmnum}[1]{\expandafter\@slowromancap\romannumeral #1@}
\makeatother

\usepackage{balance}
\usepackage[utf8]{inputenc}
\usepackage[english]{babel}

\usepackage[mathscr]{euscript}

\newtheorem{theorem}{Theorem}
\newtheorem{lemma}{Lemma}
\newtheorem{proposition}{Proposition}

\newtheorem{definition}{Definition}

\addto\captionsenglish{\renewcommand{\figurename}{Fig.}}
\usepackage[labelsep=period]{caption}

\usepackage[hidelinks]{hyperref}

\long\def\comment#1{}

\newfont{\bbb}{msbm10 scaled 700}

\newcommand{\RR}{\mathbb{ R}}

\newcommand{\EE}{\mathbb{ E}}

\newcommand{\xv}{{\bf x}}
\newcommand{\yv}{{\bf y}}

\newcommand{\zerov}{{\bf 0}}

\newcommand{\Am}{{\bf A}}

\newcommand{\Hm}{{\bf H}}
\newcommand{\Id}{{\bf I}}

\newcommand{\Lm}{{\bf L}}

\newcommand{\Sm}{{\bf S}}
\newcommand{\Tm}{{\bf T}}

\newcommand{\Vm}{{\bf V}}

\newcommand{\Zm}{{\bf Z}}

\newcommand{\Hc}{{\cal H}}

\newcommand{\Nc}{{\cal N}}

\newcommand{\Sc}{{\cal S}}

\newcommand{\Wc}{{\cal W}}
\newcommand{\Vc}{{\cal V}}

\newcommand{\RNum}[1]{\uppercase\expandafter{\romannumeral #1\relax}}

\newcommand{\sigmav}{\hbox{\boldmath$\sigma$}}

\newcommand{\Lambdam}{\hbox{\boldmath$\Lambda$}}

\newcommand{\Sigmam}{\hbox{\boldmath$\Sigma$}}

\newcommand{\trace}{{\hbox{tr}}}

\newcommand{\eqdef}{\stackrel{\Delta}{=}}

\newcommand{\squeezeequ}{\medmuskip=2mu \thinmuskip=1mu \thickmuskip=3mu}
\newcommand{\middlesqueezeequ}{\medmuskip=1.5mu \thinmuskip=0.5mu \thickmuskip=2.5mu \nulldelimiterspace=-0.5pt \scriptspace=0pt}
\newcommand{\supersqueezeequ}{\medmuskip=1mu \thinmuskip=0mu \thickmuskip=2mu \nulldelimiterspace=-1pt \scriptspace=0pt}

\newcommand{\Tsupersqueezeequ}{\medmuskip=0.1mu \thinmuskip=0mu \thickmuskip=0.1mu \nulldelimiterspace=-1pt \scriptspace=0pt}

\def\LRT#1#2{\!
\raisebox{.2ex}{$
{{\scriptstyle\;#1}\atop{\displaystyle\gtrless}}
\atop
{\raisebox{-1.25ex}{$\scriptstyle\;#2$}}
$}
\!}

\begin{document}
%
\title{Asymptotic Learning Requirements for Stealth Attacks on Linearized State Estimation}
%
%
%

\author{Ke Sun,
        I\~naki Esnaola,
        Antonia M. Tulino,
       and H. Vincent Poor

\thanks{

K. Sun was with the Department of Automatic Control and Systems
Engineering, University of Sheffield, Sheffield S1 3JD, UK
(email: cn.kesun@gmail.com).

I. Esnaola is with the Department of Automatic Control and Systems
Engineering, University of Sheffield, Sheffield S1 3JD, UK, and with the Department of Electrical and Computer Engineering, Princeton University, Princeton, NJ 08544 USA
(email: esnaola@sheffield.ac.uk).

A.~M. Tulino is with the DIETI, Universit\`{a} degli Studi di Napoli Federico II,  Naples 80138, Italy, and with the ECE Department, New York University Tandon School of Engineering, Brooklyn, NY 10003 USA
(email: antoniamaria.tulino@unina.it).

H.~V. Poor is with the Department of Electrical and Computer Engineering, Princeton
University, Princeton, NJ 08544 USA 
(e-mail: poor@princeton.edu).
}}

%



\maketitle

\begin{abstract}

Information-theoretic stealth attacks are data injection attacks that minimize the amount of information acquired by the operator about the state variables, while simultaneously limiting the Kullback-Leibler divergence between the distribution of the measurements under attack and the distribution under normal operation with the aim of controling the probability of attack detection. 
For Gaussian distributed state variables, attack construction requires knowledge of the second order statistics of the state variables, which is estimated from a finite number of past realizations using a sample covariance matrix. 
Within this framework, the attack performance is studied for the attack construction with the sample covariance matrix. 
This results in an analysis of the amount of data required to learn the covariance matrix of the state variables used on the attack construction. 
The ergodic attack performance is characterized using asymptotic random matrix theory tools, and the variance of the attack performance is bounded. 
The ergodic performance and the variance bounds are assessed with simulations on IEEE test systems.
\end{abstract}

\begin{IEEEkeywords}
Data injection attack, information-theoretic stealth attacks, statistical learning, random matrix theory, ergodic performance, variance of performance
\end{IEEEkeywords}

%
\IEEEpeerreviewmaketitle

\section{Introduction}
%
%
%
%

\IEEEPARstart{D}{ata} injection attacks (DIAs) are a type of cyber-security threat that aim to modify the  data exchange between the components of a power system by exploiting existing vulnerabilities of the sensing and communication systems \cite{giani_viking_2009}.
Specifically, DIAs alter the measurements acquired by monitoring equipment, such as remote terminal units, to corrupt state estimates without triggering bad data detection mechanism used by the operator \cite{liu_false_2009}.
Attack construction strategies vary for different estimation frameworks and abnormal data detection approaches. 
For the conventional framework consisting of least squares estimation and residual-based detection, it is shown in \cite{liu_false_2009} that attacks that lie in the column space of the system Jacobian matrix are undetectable.
However, the rapid growth of the cyber layer in the smart grid enables novel estimation and detection methods that incorporate system model knowledge and integration of multiple data sources. As a result, efficient attack strategies need to adapt in this changing landscape to multiple estimation paradigms.
For instance, within a Bayesian framework with minimum mean square error (MMSE) estimation, the attack trades off mean square error disruption for  probability of detection \cite{kosut_malicious_2011, esnaola_maximum_2016}. 
In particular, the attack construction leverages the linear MMSE estimate to identify the signal subspace that is most vulnerable to additive distortion while minimizing the probability of detection of a likelihood ratio test.

There are also multiple variants that incorporate different operational constraints for the attacker. 
For instance, introducing the $\ell_0$ norm of the attack vector  as minimization objective yields sparse attacks that decrease the number of sensors that need to be compromised by the attacker to decrease the difficulty of launching the attack \cite{dan_stealth_2010,sandberg_security_2010,sou_exact_2013,kim_strategic_2011}.
Sparse attacks are constructed in a distributed settings with multiple attackers in \cite{tajer_distributed_2011} and \cite{ ozay_sparse_2013}.
Information-theoretic attack constructions \cite{Sun_information-theoretic_2017, Sun_Stealth_2020} stem from the aspiration of targeting universal disruption measures that pose a data-integrity threat for the operator under a wide range of estimation frameworks. 
In this setting, the attacker aims to limit the amount of information acquired by the operator from the grid measurements by constructing random attacks that minimize the mutual information between the measurements and the state variables.
These attack constructions require prior knowledge about the power system, specifically the Jacobian matrix of the power system and the distribution of the state variables, to determine the distribution of the attack vectors. 
Under a Gaussian assumption for the state variables \cite{kosut_malicious_2011, esnaola_maximum_2016, Sun_Stealth_2020}, the knowledge required reduces to  the second order statistics of the state variables. 
Naturally, perfect knowledge of the second order statistics of the state variables is not attainable in practice, and as a result, the performance of such attacks degrades as a result of having imperfect statistics.
With that motivation, we study the performance loss faced by an attacker as a result of having access to imperfect knowledge about the distribution of the state variables. Specifically, we study the setting in which the attacker has access to a finite number of realizations of the state variables, for instance obtained from historical data, and learns the distribution of the state variables from them.

Imperfect system knowledge has been studied in different settings.
For example, it is shown in \cite{kim_subspace_2015} that stealthy attacks can be constructed when the attacker has only partial information about the grid.
Undetectable attack constructions that rely on imprecise branch parameters are studied in \cite{rahman_false_2012}.
Historical operation data can also be leveraged to infer the power system parameters and obtain a statistical description of the system, which can be capitalized by the attacker.
For instance, historical data is exploited to learn the topology of the grid in \cite{li_blind_2013}. 
Furthermore, different statistical learning algorithms, such as principal component analysis \cite{yu_blind_2015} and independent component analysis \cite{esmalifalak_stealthy_2018},  can successfully be used to infer the statistical distribution of the state variables and the measurements.

In this paper, we characterize the learning requirements for the information-theoretic stealth attacks proposed in \cite{Sun_information-theoretic_2017}  and \cite{Sun_Stealth_2020} via asymptotic analysis tools from random matrix theory (RMT).
To that end, we adopt a sample covariance matrix estimation framework as in \cite{sun_2019_learning}. In this setting, the attacker computes the sample covariance matrix from past realizations of the state variables and uses the estimate of the covariance matrix to construct the attack vector. Since the sample covariance estimate is asymptotically unbiased and the information-theoretic attack construction is linear in the covariance matrix \cite{Sun_information-theoretic_2017}, the attack construction resulting from using the sample covariance matrix is asymptotically optimal. 
Assuming that the historical samples form a sequence of independent and identically distributed (i.i.d.) random variables, the sample covariance matrix is a \emph{random matrix} and the performance of the attacks is a \emph{random variable}. 
The non-asymptotic performance, i.e. the case with a finite number of realizations to compute the sample covariance matrix, is studied in \cite{sun_2019_learning} but the study using RMT tools in \cite{sun_2019_learning} only provides bounds on the expected value of the performance. 
In this paper, we focus on the asymptotic regime and characterize analytically the ergodic attack performance and its variance. 
In doing so, we establish performance tradeoffs between the size of the data set used for computing the sample covariance matrix and the performance of the attacks. 
We also obtain variance bounds on the attack performance to characterize the distribution of the attack performance. 

Asymptotic RMT tools are well suited for modeling power systems with incomplete state information and provide good performance evaluations of finite dimensional systems given their rapid convergence properties \cite{vershynin_high_2018}.
RMT tools have been successfully used in the analysis of power systems before, for instance in \cite{he_big_2017, he_novel_2018, he_designing_2016}, and  \cite{cai_3d_2015} for the measurements from the IEEE test systems. 
Specifically, \cite{he_big_2017} and \cite{he_novel_2018} show that the distribution of the singular values or eigenvalues of voltage data from IEEE test systems satisfies the Mar\u{c}enko–Pastur law and the circular law, which are used in \cite{he_designing_2016} to detect abnormal events and in \cite{cai_3d_2015} for data visualization. 
In a cybersecurity context, RMT tools are used in \cite{lakshminarayana_2020_data} to characterize the tradeoff between the sparsity of the attacks and the probability of passing the detection mechanism when a limited number of measurements are available for the attacker.

The rest of the paper is organized as follows: 
In Section \ref{System_Model}, a Bayesian framework with linearized dynamics for DIAs is presented.
The learning scenario and some auxiliary asymptotic RMT results are presented in Section \ref{Sec:setting}. 
Using these results, a closed-form expression for the ergodic attack performance and bounds for the variance of the performance are proposed in Section \ref{Sec:Ergodic} and Section \ref{Sec:variance}, respectively, for the attack constructed using imperfect information.
Section \ref{Sec:Numerical_Simulation} numerically evaluates the results in Section \ref{Sec:Ergodic} and Section \ref{Sec:variance} on IEEE test systems.
The paper ends with conclusions in Section \ref{Sec:Conclusion}.

\section{System Model} \label{System_Model}

\subsection{Bayesian Framework with Linearized Dynamics}
The measurement model for state estimation with linearized dynamics is given by
\begin{IEEEeqnarray}{c}\label{Equ:DCSE}
Y^{m} = \Hm X^{n} + Z^{m},
\end{IEEEeqnarray}
where $Y^{m} \in \RR^{m}$ is a vector of random variables describing the measurements; $X^{n} \in \RR^{n}$ is a vector of random variables describing the state variables;
$\Hm \in \RR^{m\times n}$ is the linearized Jacobian measurement matrix that is given by 
 \begin{IEEEeqnarray}{c}\label{Equ:Hm}
\Hm = \frac{\partial H(X^n)}{ \partial X^{n}}\Big|_{X^{n} = \xv_0}, 
\end{IEEEeqnarray}
in which $H: \RR^{n} \to \RR^{m}$ models  the nonlinear dynamics between the state variables and the measurements and $\xv_0$ is the operating point; 
and $Z^{m} \in \RR^{m} $ is the additive white Gaussian noise (AWGN) with distribution $\Nc(\zerov,\sigma^{2} \Id)$ that is introduced by the sensors as a result of the thermal noise, in which $\Id$ denotes the identity matrix of proper dimension and $\sigma^2$ is the variance of the noise  \cite{abur_power_2004, grainger_power_1994} {\footnote{ Note that the linearized model is not limited to the direct current (DC) model in the power system state estimation. 
This model allows the operator to include both bus voltage magnitudes and angles as state variables, such as PMU-based state estimation case, in the linearized model, see \cite[Table 15.4, Table 15.5]{grainger_power_1994}.
}.

In the remaining of the paper, we assume that the vector of state variables follows a zero-mean multivariate Gaussian distribution given by
\begin{IEEEeqnarray}{c}\label{Equ:SV_Gaussian}
  X^{n} \sim \Nc(\zerov,\Sigmam_{X\!X}),
\end{IEEEeqnarray}
where $\Sigmam_{X\!X}\in \Sc^{n}_{+}$ is the covariance matrix of the distribution of the state variables and $\Sc^{n}_{+}$ denotes the set of positive semidefinite matrices of size $n\times n$. 
As a result of the linearized model in (\ref{Equ:DCSE}), the vector of measurements also follows a multivariate Gaussian distribution denoted by
\begin{IEEEeqnarray}{c}\label{Equ:measurement_dis}
  Y^{m} \sim \Nc(\zerov,\Sigmam_{Y\!Y}),
\end{IEEEeqnarray}
where $\Sigmam_{Y\!Y} = \Hm\Sigmam_{X\!X}\Hm^{\sf T} + \sigma^{2}\Id$ is the relative covariance matrix.

The formulation of the problem in a Bayesian setting demands the introduction of a modeling assumption on the distribution of the state variables. 
The rationale for adopting a Gaussian distribution over the state variables in this paper stems from a maximum entropy \cite{jaynes_1957_information} approach.
Indeed, the distribution that maximizes the entropy for given second order moments is the Gaussian distribution \cite{cover_elements_2012}. 
As a result, by adopting a Gaussian model, we introduce the minimum amount of bias into our modeling, i.e. we adopt the distribution that is maximally non-committal over the state variables. 
From a modeling perspective, this is the distribution that assumes the least amount of prior over the problem formulation given that we only have information about the second order statistics. 
Gaussian models have been previously used in power flow problems 
\cite{constante_2018_data, schellenberg_2005_cumulant} and DIAs \cite{kosut_malicious_2011, esnaola_maximum_2016, Sun_Stealth_2020}. 
Interestingly, Gaussian distributions have also been observed in some of the smart grid processes. 
For instance, real data suggests that distribution networks are well described by Gaussian models for both state variables \cite{genes_recovering_2016} and consumption measurements \cite{woolley_statistical_2012}. 
This suggests that nodal active power injections can be modeled as Gaussian but the fitness of the model to other state variables, such as reactive power or PMU measurements, requires further study. 
It is worth noting, therefore, that the insight provided by the analytical results in this paper is more reliable the closer the state variables are to following a Gaussian distribution. 
In any case, the linear observation model in \eqref{Equ:DCSE} gives rise to Gaussian distributions over the state variables whenever the observations follow a Gaussian distribution or when the deviations with respect to the operation point of the state variables can be modeled by Gaussian distributions. 

%

Under our setting, DIAs corrupt the measurements available to the operator by adding an attack vector of random variables to the measurements.
The resulting vector of compromised measurements is given by
\begin{IEEEeqnarray}{c}
\label{eq:measurement_model}
Y^{m}_{A} = \Hm X^{n} + Z^{m} + A^{m},
\end{IEEEeqnarray}
where $A^{m} \in \RR^{m}$ is the attack vector and $Y^{m}_{A} \in \RR^{m } $ is the vector containing the compromised measurements \cite{liu_false_2009}.
Following the setting of \cite{Sun_information-theoretic_2017} and \cite{Sun_Stealth_2020}, 
an attack vector which is independent of the state variables is constructed under the assumption that the attack vector follows a multivariate Gaussian distribution denoted by
\begin{IEEEeqnarray}{c}
  A^{m} \sim  \Nc (\zerov,\Sigmam_{A\!A}), 
\end{IEEEeqnarray}
where $\Sigmam_{A\!A}\in \Sc^{m}_{+}$ is the associated covariance matrix.
Because of the Gaussianity of the attack distribution, the vector of compromised measurements is distributed as
\begin{IEEEeqnarray}{c}\label{Equ:Dis_Ya}
  Y_{A}^{m} \sim \Nc(\zerov,\Sigmam_{Y_{A}\!Y_{A}}),
\end{IEEEeqnarray}
where $\Sigmam_{Y_{A}\!Y_{A}} = \Hm\Sigmam_{X\!X}\Hm^{\sf T} + \sigma^{2}\Id + \Sigmam_{A\!A} $. 

The operator of the power system makes use of the acquired measurements to detect the attack.
The detection problem is cast as a hypothesis testing problem with hypotheses
\begin{IEEEeqnarray}{cl}
\Hc_{0}:  \ & Y^{m} \sim \Nc(\zerov,\Sigmam_{Y\!Y}), \quad \text{versus} \label{Equ:HT_0} \\
\Hc_{1}:  \ & Y^{m} \sim \Nc(\zerov,\Sigmam_{Y_{A}\!Y_{A}}). \label{Equ:HT_1} 
\end{IEEEeqnarray}
The null hypothesis $\Hc_{0}$ describes the case in which the power system is not compromised, while the alternative hypothesis $\Hc_{1}$ describes the case in which the power system is under attack.
Note that by assuming that the operator knows the distribution of the attack vector, the attacker aims to induce a probability of detection that is close to the probability of false alarm. 
When the probability of false alarm is comparable to the probability of attack detection, the operator is unable to distinguish between bad data events during normal operation and bad data resulting from DIAs.


For the binary hypothesis testing problem in (\ref{Equ:HT_0}) and (\ref{Equ:HT_1}),  Neyman-Pearson lemma states that likelihood ratio test (LRT) is the most powerful test under a prefixed significance level $\alpha$, i.e.  LRT achieves the maximum probability of detection among all the tests with probability of false alarm smaller that $\alpha$ \cite[Proposition \RNum{2}.D.1]{poor_introduction_1994}. 
As a result, the LRT is used to decide between $\Hc_{0}$ and $\Hc_{1}$ based on the available measurements.
The LRT between $\mathcal{H}_{0}$ and $\mathcal{H}_{1}$ takes following form:
\begin{equation}\label{LHRT}
L(\yv) \eqdef \frac{f_{Y^{m}_{A}}(\yv)}{f_{Y^{m}}(\yv)} \ \LRT{\Hc_{1}}{\Hc_{0}} \ \tau,
\end{equation}
where $\yv \in \RR^{m}$ is a realization of the vector of random variables modeling the measurements; $f_{Y_A^m}$ and  $f_{Y^m}$ denote the probability density functions of $Y_A^m$ and  $Y^m$, respectively; and $\tau$ is the decision threshold set by the operator to meet a given false alarm constraint.

\subsection{Information-Theoretic Stealth Attacks}\label{Subsec:Information_Theoretic_setting}

The probabilistic modeling of the system variables enables an information-theoretic analysis of the DIAs \cite{Sun_information-theoretic_2017,Sun_Stealth_2020}.
The measurement model in (\ref{Equ:DCSE}) characterizes an information acquisition procedure, in which the operator acquires the information about the state variables from the gathered measurements. 
To that end, the attacker disrupts the information acquisition procedure by minimizing the amount of information acquired by the operator, or mathematically, by minimizing the mutual information between the vector of state variables and the vector of compromised measurements, i.e. minimizing $I(X^{n};Y_{A}^{m})$ in (\ref{eq:measurement_model}), where $I(\cdot \hspace{0.1em};\cdot)$ denotes the mutual information.
Given that the smart grid paradigm envisions a large array of data-driven processes taking place in the system, the use of mutual information as the measure of the utility of the data is reasonable given the fundamental character of the mutual information. 
Indeed, the links of mutual information to detection \cite{vazquez_2016_bayesian}, estimation theory \cite{guo_2013_interplay}, and machine learning \cite{russo_2019_much} problems facilitate results in an attack disruption metric with operational meaning on a wider range of applications.

From the perspective of the attacker, the attacker also needs to guarantee the attacks to be stealthy under the detection approach, which requires  the minimization of the probability of detection under the detection approach. 
In particular, minimizing the probability of detection under the LRT in (\ref{LHRT}) is achieved by minimizing the asymptotic value of the probability,  and as a result of the Chernoff-Stein lemma \cite[Theorem 11.7.3]{cover_elements_2012} \cite[(10)]{Sun_Stealth_2020}, it is equivalent to minimizing $D(P_{Y^{m}_{A}}\|P_{Y^{m}})$, where $P_{Y_A^m}$ and  $P_{Y^m}$ denote the probability distributions of $Y_A^m$ and  $Y^m$, respectively, and $D(\cdot \|\cdot)$ denotes the Kullback-Leibler (KL) divergence.


The \emph{stealth attacks} minimize the amount of information acquired by the operator and the probability of attack detection simultaneously by minimizing the \textit{data integrity} cost function given by 
\begin{IEEEeqnarray}{c}\label{Equ:Stealth_obj}
\underset{A^{m}}{\text{min}} \ I\left(X^{n};Y^{m}_{A}\right)  +   D\left( P_{{Y}^{m}_{A}}\|P_{Y^{m}} \right),
\end{IEEEeqnarray}
which is equivalent to 
\begin{IEEEeqnarray}{c}
\underset{A^{m}}{\text{min}} \ D( P_{X^{n}Y_{A}^{m}}\|P_{X^{n}}P_{Y^{m}})\label{Equ:Stealth_obj_1}
\end{IEEEeqnarray}
after some algebraic operations \cite{hou_effective_2014}, where $P_{X^{n}Y_{A}^{m}}$ is the joint distribution of $(X^{n}, Y_{A}^{m})$.
The unweighted sum in (\ref{Equ:Stealth_obj}) is generalized into a weighted sum in \cite{Sun_Stealth_2020}.
Under the Gaussian assumption for the state variables and the attack, it is shown in \cite{Sun_information-theoretic_2017} that the data integrity cost function in (\ref{Equ:Stealth_obj_1}) is a convex function of the attack covariance matrix $\Sigmam_{A\!A}$ with optimal solution
\begin{IEEEeqnarray}{c}\label{Equ:Stealth_optimal}
 \Sigmam_{A\!A}^{\star} = \Hm\Sigmam_{X\!X}\Hm^{\sf T}.
\end{IEEEeqnarray}

A detailed account of information-theoretic stealth attacks and the tradeoff between disruption and probability of detection are provided in \cite{Sun_information-theoretic_2017}, \cite{Sun_Stealth_2020}, and \cite{esnaola_data_2021}.

\section{Learning Scenario for Stealth Attacks}\label{Sec:setting}

\subsection{Learning Scenario Setting}\label{Sec:LSS}
As shown in (\ref{Equ:Stealth_optimal}), the attacker needs to get access to the system Jacobian matrix $\Hm$ and the covariance matrix of the state variables $\Sigmam_{X\!X}$.
Note that the setting in this paper differs from the setting in \cite{liu_false_2009} in that the attacker exploits knowledge of the second order statistics of the state variables. 
To that end, the attacker estimates the covariance matrix $\Sigmam_{X\!X}$ based on the available training data. 
The amount of training data governs the accuracy of the covariance matrix estimate, and therefore, in practical settings in which the attacker has access to a finite number of historical data samples, the attack is constructed with partial knowledge of the covariance matrix.

%

In the following, we study the performance of the attack when the covariance matrix is not perfectly known by the attacker but the linearized Jacobian measurement matrix is known.
We model the partial knowledge by assuming that the attacker has access to a sample covariance matrix of the state variables.
Specifically, a training dataset consisting of an i.i.d. sequence of $k$ samples of the state variables, i.e. of $\{X^n_{i}\}^{k}_{i=1}$, is available to the attacker.
This assumes that the attacker has access to noiseless realizations of the state variables that can be used to estimate the state variables without any error. While in practical settings access to noiseless realizations is not feasible, this assumption aims to model the worst-case scenario attack for the operators, i.e. the case in which the attacker has access to perfect historical data of the state variables.

That being the case, the attacker computes the unbiased estimate of the mean and the covariance matrix of the state variables via
\begin{IEEEeqnarray}{c}
k \bar{X}^n = \sum_{i=1}^{k} X^n_{i}, \\
 \ (k-1)\Sm_{X\!X} = \sum_{i=1}^{k} X^n_{i} (X^n_{i})^{\sf T} - k \bar{X}^n(\bar{X}^n)^{\sf T}, \label{Equ:SC}\supersqueezeequ \IEEEeqnarraynumspace
\end{IEEEeqnarray}
where $\bar{X}^n$ is the sample mean and $\Sm_{X\!X}$ is the sample covariance matrix. 
Given that the vector of the state variables follows a multivariate normal distribution, it is shown in \cite[Proposition 7.1]{bilodeau_2008_theory} that the sample covariance matrix in (\ref{Equ:SC}) is a random matrix with a central Wishart distribution given by 
\begin{IEEEeqnarray}{c}\label{Equ:Wishart_S_XX}
(k-1)\Sm_{X\!X} \sim \Wc_{n} (k-1, \Sigmam_{X\!X}),
\end{IEEEeqnarray}
where $\Wc_{n} (k-1, \Sigmam_{X\!X})$ denotes the central Wishart distribution with $k-1$ degrees of freedom and covariance matrix $\Sigmam_{X\!X}$.

Given the optimal stealth attacks expression in (\ref{Equ:Stealth_optimal}), the stealth attacks constructed using the sample covariance matrix follow a multivariate Gaussian distribution conditioned on the sample covariance matrix $\Sm_{X\!X}$, which is given by
\begin{IEEEeqnarray}{c} \label{Equ:tilde_A}
\tilde{A}^{m} \sim \Nc (\zerov, \Sigmam_{\tilde{A}\!\tilde{A}})
\end{IEEEeqnarray}
with $\Sigmam_{\tilde{A}\!\tilde{A}} = \Hm\Sm_{X\!X}\Hm^{\sf T}$; and as a result of (\ref{Equ:Wishart_S_XX}) and \cite[Proposition 7.4]{bilodeau_2008_theory}, it holds that
\begin{IEEEeqnarray}{c}\label{Equ:Wishart_S_AA}
(k-1)\Sigmam_{\tilde{A}\!\tilde{A}} = (k-1)\Hm\Sm_{X\!X}\Hm^{\sf T} \sim \Wc_{m} (k-1, \Hm \Sigmam_{X\!X}\Hm^{\sf T} ). \supersqueezeequ \IEEEeqnarraynumspace
\end{IEEEeqnarray}
To that end, the measurements that are under attack are given by
\begin{IEEEeqnarray}{c}
Y_{\tilde{A}}^{m} = \Hm X^{n} + Z^{m} + \tilde{A}^{m},
\end{IEEEeqnarray}
in which $Y_{\tilde{A}}^{m} \in \RR^{m} $  is the vector containing the measurements that are compromised by the attacks in (\ref{Equ:tilde_A}). 
As a result, the compromised measurements follow a multivariate Gaussian distribution conditioned on the sample covariance matrix $\Sm_{X\!X}$, i.e. 
\begin{IEEEeqnarray}{c}
Y_{\tilde{A}}^{m} \sim \Nc (\zerov, \Sigmam_{Y_{\tilde{A}}\!Y_{\tilde{A}}})
\end{IEEEeqnarray}
with $\Sigmam_{Y_{\tilde{A}}\!Y_{\tilde{A}}} = \Hm\Sigmam_{X\!X}\Hm^{\sf T} + \sigma^{2}\Id + \Sigmam_{\tilde{A}\!\tilde{A}}$. 
Similarly, the cost function in (\ref{Equ:Stealth_obj_1}) is described in this case as
\begin{equation}
\label{eq:ergodic_cost}
D\left( P_{X^{n}Y_{\tilde{A}}^{m}}\|P_{X^{n}}P_{Y^{m}}\right),
\end{equation}
where $P_{X^{n}Y_{\tilde{A}}^{m}}$ is the joint distribution of $(X^{n}, Y_{\tilde{A}}^{m})$. 
Under the Gaussianity assumption, (\ref{eq:ergodic_cost}) is equivalent to \cite[Proposition 1]{Sun_Stealth_2020}
\begin{IEEEeqnarray}{c} \label{Equ:obj_rv_0}
F \left(\Sigmam_{\tilde{A}\!\tilde{A}}\right)   \eqdef  \ \frac{1}{2}  \left(\trace(\Sigmam_{Y\!Y}^{-\!1}\Sigmam_{\tilde{A}\!\tilde{A}})- \log |\Sigmam_{\tilde{A}\!\tilde{A}}+\sigma^{2}\Id|+\log|\Sigmam_{Y\!Y}|\right) .\IEEEeqnarraynumspace \supersqueezeequ
\end{IEEEeqnarray}

Given the Wishart distribution of the attack covariance matrix in (\ref{Equ:Wishart_S_AA}), the KL divergence objective in (\ref{eq:ergodic_cost}) and the cost functions in (\ref{Equ:obj_rv_0}) are both \emph{random variables}.
Following on the same steps as in \cite{sun_2019_learning}, we defined the \emph{ergodic performance} of the attack as the performance obtained by averaging over all realizations of the training data set, i.e. as $\EE\!\left[ F \!\left(\Sigmam_{\tilde{A}\!\tilde{A}}\right) \right]$.

The characterization of the objective and the cost function using RMT can be conducted in the non-asymptotic scenario and the asymptotic scenario.
Note that under both scenarios the characterization that uses the distribution of the Wishart random matrices directly is not manageable, so we turn to the distribution of the eigenvalues of Wishart random matrices\footnote{The expression in right-hand side of (\ref{Equ:obj_rv_0}) can be rewritten as a function of the eigenvalues of $\Sigmam_{\tilde{A}\!\tilde{A}}$. We will show this later in Theorem \ref{Pro_EP}.}, which is more tractable. 
The non-asymptotic scenario focuses on the case when the dimensions of the random matrices are finite value, i.e. $k-1$ and $n$ are finite integers.
Under this scenario, only probabilistic bounds are available for the eigenvalues of random matrices. 
To that end, we can only provide upper and lower bounds on the non-asymptotic ergodic performance \cite{sun_2019_learning}.

Unlike the non-asymptotic scenario, the asymptotic scenario focuses on the case when the dimension of the random matrices goes to infinite, i.e. when $k-1 \to \infty$ and $n \to \infty$.
The rapid convergence of the non-asymtotic results to the asymptotic results guarantees that the asymptotic results approximate the non-asymptotic results well even for small values of $k-1$ and $n$, which will be shown later by the numerical simulation result in Fig. \ref{Fig:ST_Rho_30_SNR30}. 
The eigenvalue distributions that arise for these types of matrices in the asymptotic case are simpler to describe, and therefore, amenable to be studied analytically.  
In fact for this case, we are able to obtain a closed-form expression for the ergodic performance, rather than bounds as in the non-asymptotic case \cite{sun_2019_learning}.
Also the variance of the performance, i.e. $\textnormal{var}\left[ F \left(\Sigmam_{\tilde{A}\!\tilde{A}}\right) \right]$, can also be characterized by the corresponding bounds. 

In the following, we analyze the performance of the attack constructed using the sample covariance matrix for the asymptotic scenario.
Before that, we introduce some auxiliary asymptotic results from RMT to aid the analysis.

\subsection{Auxiliary Results from RMT}

Asymptotic RMT mainly investigates the spectral properties of random matrices when the dimension tends to infinity \cite{bai_spectral_2010}.
As the dimension increases to infinity, the distribution of the eigenvalues of random matrices converges to the fixed distributions, such as the Mar\u{c}enko–Pastur law for Wishart random matrices.
We first introduce some definitions from RMT. 

\begin{definition} \cite{tulino_random_2004}\label{Def:eat}
 The $\eta$-transform of a nonnegative random variable $X$ is
\begin{align}
\eta_{X}(\gamma) = \EE \left[ \frac{1}{1+\gamma X}\right],
\end{align}
where $\gamma \geq 0$ and thus $1 \geq\eta_{X}(\gamma) >0$.
\end{definition}

\begin{definition} \cite{tulino_random_2004}\label{Def:ST}
The Shannon transform of a nonnegative random variable $X$ is defined as
\begin{align}\label{Equ:ST_DEF}
\Vc_{X}(\gamma) = \EE \left[ \log(1+\gamma X)\right].
\end{align}
\end{definition}

\begin{definition} \cite{tulino_random_2004}
The asymptotic eigenvalue distribution (AED), $\textnormal{\textsf{F}}_{\mathbf{A}} (\cdot)$, of an $n \times n$ Hermitian random matrix $\Am$ is defined as
\begin{align}
\textnormal{\textsf{F}}_{\mathbf{A}} (x) = \lim_{n \to \infty} \frac{1}{n} \sum_{i=1}^{n} \mathbbm{1}_{\left\{\lambda_{i}(\mathbf{A}) \leq x \right\}},
\end{align}
where $\mathbbm{1}_{\{\cdot\}}$ is the indicator function and $\lambda_{1}(\mathbf{A}), \dots, \lambda_{n}(\mathbf{A})$ are the eigenvalues{\footnote{The eigenvalues $\lambda_1, \ldots, \lambda_n$ are unordered eigenvalues.} of $\Am$}.
\end{definition}

The following theorems characterize the Shannon transform of the spectral distribution of a certain type of random matrices and the variance of the logarithm of the spectral distribution. 
These results enable us to characterize the asymptotic case better. 

\begin{theorem}\label{Theorem:ST}\cite[Theorem 2.39]{tulino_random_2004} 
Let $\Lm$ be an $n\times (k-1)$ matrix whose entries are zero-mean i.i.d. random variables with variance $\frac{1}{k-1}$. 
Let $\Tm$ be an $n \times n$ symmetric nonnegative random matrix, independent of $\Lm$, whose AED converges almost surely to a nonrandom limit. 
The AED of  $\Lm^{\sf T} \Tm \Lm$ converges almost surely, as $k-1, n \to \infty$ with $\frac{k-1}{n} \to \beta$, to a distribution whose $\eta$-transform satisfies 
\begin{align}\label{Equ:ST_HTH_1}
\frac{1}{\beta} = \frac{1-\eta}{1-\eta_{\mathbf{T}}(\gamma \eta)},
\end{align}
where for notational simplicity we have abbreviated $\eta_{\mathbf{L}^{\sf T} \Tm \mathbf{L}}(\gamma) = \eta$.
The corresponding Shannon transform satisfies 
\begin{align} \label{Equ:ST_1}
\Vc_{\mathbf{L}^{\sf T} \Tm \mathbf{L}} (\gamma ) = \frac{\Vc_{ \mathbf{T}}(\gamma \eta) }{\beta} + \log \frac{1}{\eta} + (\eta-1) \log e.
\end{align}
\end{theorem}

Note that the definition of $\beta$ here is the reciprocal of the definition in \cite{tulino_random_2004}. 

\begin{theorem} \cite[Theorem 4]{tulino_asymptotic_2005} \label{Theorem:ST_Variance}
Let $\Lm$ be an $n\times (k-1)$ matrix defined as in Theorem \ref{Theorem:ST}. 
Let $\Tm$ be an $n \times n$ matrix defined as in Theorem \ref{Theorem:ST} whose the spectral norm is bounded.
As $k-1,n \to \infty$ with $\frac{k-1}{n} \to \beta$, the random variable 
\begin{IEEEeqnarray}{c}\label{Equ:Variance_inte_3}
\Delta_{k-1} = \log |\Id + \gamma \mathbf{L}^{\sf T} \Tm \mathbf{L} | - (k-1) \Vc_{\mathbf{L}^{\sf T} \Tm \mathbf{L}} (\gamma) \IEEEeqnarraynumspace
\end{IEEEeqnarray}
is asymptotically zero-mean Gaussian with variance 
\begin{IEEEeqnarray}{c}\label{Equ:var_logdet}
\EE \left[\Delta^2\right] = -\log \left( 1 - \frac{1}{\beta} \EE \left[ \left( \frac{T\gamma \eta_{\mathbf{L}^{\sf T} \Tm \mathbf{L}}(\gamma)}{1+T\gamma \eta_{\mathbf{L}^{\sf T} \Tm \mathbf{L}}(\gamma)}\right)\right]\right), \IEEEeqnarraynumspace
\end{IEEEeqnarray}
where the expectation is over the nonnegative random variable $T$, whose distribution is the AED of $\Tm$.
\end{theorem}

To obtain the result in Theorem \ref{Theorem:ST_Variance}, a central limit theorem result is needed for the linear spectral statistics of random matrices.
We introduce it in the following theorem.

\begin{theorem} \cite{tulino_asymptotic_2005, bai_2004_clt} \label{Theorem_Variance}
Let $\Lm$ be an $n\times (k-1)$ matrix defined as in Theorem \ref{Theorem:ST}. 
Let $\Tm$ be an $n \times n$ matrix defined as in Theorem \ref{Theorem:ST} whose the spectral norm is bounded. 
Let $g(\cdot)$ be a continuous function on the real line with bounded and continuous derivatives, analytic on an open set containing the interval 
\begin{IEEEeqnarray}{c}
\left[ \underset{n}{\textnormal{lim inf}} \ \phi_{n} \textnormal{max} \left\{ 0, 1-\sqrt{1/\beta}\right\}^2, \  \underset{n}{\textnormal{lim sup}} \ \phi_{1} ( 1+\sqrt{1/\beta})^2\right] 
\IEEEeqnarraynumspace \supersqueezeequ
\end{IEEEeqnarray}
where $\phi_1 \geq \dots \geq \phi_n$ are the eigenvalues of $\Tm$.
Denoting the $i$-th eigenvalue and asymptotic AED of $\mathbf{L}^{\sf T} \Tm \mathbf{L}$ by $\lambda_i(\mathbf{L}^{\sf T} \Tm \mathbf{L})$ and $\textnormal{\sf F}_{\mathbf{L}^{\sf T} \Tm \mathbf{L}}(\cdot)$, the random variable
\begin{IEEEeqnarray}{c}
\Delta_{k-1} = \sum_{i=1}^{k-1} g\left(\lambda_i(\mathbf{L}^{\sf T} \Tm \mathbf{L})\right) - (k-1) \int g(x) \textnormal{d}\textnormal{\sf F}_{\mathbf{L}^{\sf T} \Tm \mathbf{L}} \IEEEeqnarraynumspace
\end{IEEEeqnarray}
converges, as $k-1,n \to \infty$ with $\frac{k-1}{n} \to \beta$, to a zero-mean Gaussian random variable with variance 
\begin{IEEEeqnarray}{c}\label{Equ:Variance_inte_1}
\EE \left[\Delta^2\right] = -\frac{1}{2\pi^2} \oint \oint \frac{\dot{g} (\mathscr{Z}(u_1))g (\mathscr{Z}(u_2))}{u_2 - u_1} \textnormal{d} u_1  \textnormal{d} u_2 \IEEEeqnarraynumspace
\end{IEEEeqnarray}
or 
\begin{IEEEeqnarray}{c}\label{Equ:Variance_inte_2}
\EE \left[\Delta^2\right] = -\frac{1}{2\pi^2} \oint \oint \frac{g (\mathscr{Z}(u_1))g (\mathscr{Z}(u_2))}{(u_2 - u_1)^2} \textnormal{d} u_1  \textnormal{d} u_2, \IEEEeqnarraynumspace
\end{IEEEeqnarray}
where $\dot{g}(x) = \frac{\textnormal{d}}{\textnormal{d}x} g(x)$ while 
\begin{IEEEeqnarray}{c}\label{Equ:Variance_z}
\mathscr{Z}(u) = - \frac{1}{u} \left( 1-\frac{1 - \eta_{\mathbf{T}}(u)}{\beta}\right).
\end{IEEEeqnarray}
In (\ref{Equ:Variance_inte_1}) and (\ref{Equ:Variance_inte_2}), the integration variables $u_1$ and $u_2$ follow closed contours, which we may take to be non-overlapping and conterclockwise, such that the corresponding contours mapped through $\mathscr{Z}(u) $ enclose the support of $\textnormal{\sf F}_{\mathbf{L}^{\sf T} \Tm \mathbf{L}}(\cdot)$.
\end{theorem}

Here (\ref{Equ:Variance_inte_1}) follows from \cite[Theorem 3]{tulino_asymptotic_2005} and (\ref{Equ:Variance_inte_2}) follows from  \cite[Theorem 1.1]{bai_2004_clt}.
These two expressions are equivalent. 
The difference is that (\ref{Equ:Variance_inte_1}) is suitable for some logarithm functions, such as $g(x) = \log(1+\gamma x)$ that is used for Theorem \ref{Theorem:ST_Variance} and (\ref{Equ:Variance_inte_2}) is suitable for some linear functions, such as $g(x) = x$ that we will discuss later in Theorem \ref{Theorem:var_trace}.

It is worth mentioning that the results in Theorem \ref{Theorem:ST}, \ref{Theorem:ST_Variance}, and \ref{Theorem_Variance}, including Theorem \ref{Theorem:var_trace} in the following section, are general results, in which the distribution of entries in $\Lm$ is not specified.
The only requirement is that the matrix $\Lm$ is composed of zero-mean i.i.d. entries with normalized variance.

\subsection{Asymptotic Results for Trace Terms}

Using the result in Theorem \ref{Theorem_Variance}, we introduce an analytical expression for the variance of $\textnormal{\trace}\left(\Lm^{\sf T} \Tm \Lm \right)$, which we use later in the performance analysis. 

\begin{theorem} \label{Theorem:var_trace}
Let $\Lm$ be an $n\times (k-1)$ matrix defined as in Theorem \ref{Theorem:ST}.
Let $\Tm$ be a symmetric nonnegative definite random matrix independent of $\Zm$ with bounded spectral norm and whose asymptotic AED converges almost surely to a nonrandom limit.
As $k-1,n \to \infty$ with $\frac{k-1}{n} \to \beta$, the random variable 
\begin{IEEEeqnarray}{c}\label{Equ:Variance_inte_0}
\Delta_{k-1} = \textnormal{\trace}\left(\Lm^{\sf T} \Tm \Lm \right)  - (k-1) \frac{\EE\left[ T\right]}{\beta}
\end{IEEEeqnarray}
is asymptotically zero-mean Gaussian with variance 
\begin{IEEEeqnarray}{c}\label{Equ:var_trace}
\EE \left[\Delta^2\right] = \frac{2}{\beta} \EE\left[ T^2\right], 
\end{IEEEeqnarray}
where the distribution of $T$ is the AED of $\Tm$.
\end{theorem}

\begin{proof}
Firstly, we note that the mean of $\trace\left(\Lm^{\sf T} \Tm \Lm \right)$ is given by
\begin{IEEEeqnarray}{c}
\EE \left[\trace\left(\Lm^{\sf T} \Tm \Lm \right) \right] = \EE \left[\trace\left(\Tm \Lm \Lm^{\sf T} \right) \right] =  \EE \left[\trace\left(\Tm\right) \right] \to  n \EE \left[ T \right], \IEEEeqnarraynumspace \squeezeequ \label{Equ:var_trace}
\end{IEEEeqnarray}
where the first equality follows from the cyclic permutation property of the trace operator; 
the second equality follows from the fact that $\Tm$ is independent of $\Lm$, the trace is a linear operator, and $\EE[\Lm \Lm^{\sf T}] = \Id $.

Secondly, we turn to obtain the variance. 
Taking $g(x) = x$ into (\ref{Equ:Variance_inte_2}) yields
 \begin{IEEEeqnarray}{c}
\EE \left[\Delta^2\right] = -\frac{1}{2\pi^2} \oint \oint \frac{\mathscr{Z}(u_1)\mathscr{Z}(u_2)}{(u_2 - u_1)^2} \textnormal{d} u_1  \textnormal{d} u_2 \IEEEeqnarraynumspace
\end{IEEEeqnarray}
with $\mathscr{Z}(u)$ in (\ref{Equ:Variance_z}).
Without loss of generality, we assume that, besides satisfying the condition of Theorem \ref{Theorem_Variance}, the $u_1$ and $u_2$ contours do not overlap and that the $u_2$ contour encloses the $u_1$ contour. 
Consequently, $\EE \left[\Delta^2\right] $ is rewritten as 
 \begin{IEEEeqnarray}{c} \label{Equ:trace_proof_1}
\EE \left[\Delta^2\right] = -\frac{1}{2\pi^2} \oint \mathscr{Z}(u_2) \left[\oint \frac{\mathscr{Z}(u_1)}{(u_2 - u_1)^2} \textnormal{d} u_1 \right] \textnormal{d} u_2.  \IEEEeqnarraynumspace
\end{IEEEeqnarray}
To calculate the inner integral, we need to use Cauchy's residue theorem.  
The first step is to find the zeros and poles of $\mathscr{Z}(u_1)$. 
Finding the zeros of $\mathscr{Z}(u_1)$ is equivalent to solving 
 \begin{IEEEeqnarray}{c}
- \frac{1}{u} \left( 1-\frac{1 - \eta_{\mathbf{T}}(u)}{\beta}\right) = 0.
\end{IEEEeqnarray}
Note that, for the case $\gamma = 1$, (\ref{Equ:ST_HTH_1}) can be rewritten as 
 \begin{IEEEeqnarray}{c}
 1-\frac{1 - \eta_{\mathbf{T}}(u)}{\beta}= \eta.
\end{IEEEeqnarray}
Given the fact that $\eta \in (0,1]$, as Definition \ref{Def:eat}, so $\mathscr{Z}(u_1)$ has no zeros. 
Without lost of generality, the $u_1$ contour can be chosen such that only the simple pole at $u_1 = 0$ is enclosed. 
As a result, the inner integral is calculated using Cauchy's residue theorem and is given by 
 \begin{IEEEeqnarray}{c}
\oint \frac{\mathscr{Z}(u_1)}{(u_2 - u_1)^2} \textnormal{d} u_1  =  2\pi i \frac{-1}{u_2^2}.
\end{IEEEeqnarray}
Taking the value of the inner integral into (\ref{Equ:trace_proof_1}) yields
 \begin{IEEEeqnarray}{c}\label{Equ:trace_proof_3}
\EE \left[\Delta^2\right] = -\frac{1}{\pi i } \oint \left(- \frac{1}{u_2} \left( 1-\frac{1 - \eta_{\mathbf{T}}(u_2)}{\beta} \right)\right)   \frac{1}{u_2^2}\textnormal{d} u_2. \IEEEeqnarraynumspace
\end{IEEEeqnarray}
The proof is completed by applying Cauchy's residue theorem again for (\ref{Equ:trace_proof_3}), in which $u_2 = 0$ is a pole of order $3$.
\end{proof}

\section{Explicit Expression for the Asymptotic Ergodic Data Integrity}\label{Sec:Ergodic}

As shown in (\ref{Equ:obj_rv_0}), the objective function of the stealth attacks constructed using the sample covariance matrix is given by 
\begin{IEEEeqnarray}{c} \label{Equ:obj_rv}
F \eqdef F\!  \left(\Sigmam_{\tilde{A}\!\tilde{A}}\right)  =  \hspace{0.1em}\frac{1}{2}  \left(\trace(\Sigmam_{Y\!Y}^{-\!1}\Sigmam_{\tilde{A}\!\tilde{A}})- \log |\Sigmam_{\tilde{A}\!\tilde{A}}+\sigma^{2}\Id|+\log|\Sigmam_{Y\!Y}|\!\right), \IEEEeqnarraynumspace \Tsupersqueezeequ
\end{IEEEeqnarray}
where
\begin{IEEEeqnarray}{c}
(k-1)\Sigmam_{\tilde{A}\!\tilde{A}} = (k-1)\Hm\Sm_{X\!X}\Hm^{\sf T} \sim \Wc_{m} (k-1, \Hm\Sigmam_{X\!X} \Hm^{\sf T}). \supersqueezeequ \IEEEeqnarraynumspace
\end{IEEEeqnarray}
Note that the objective given in (\ref{Equ:obj_rv}) is a \emph{random variable}.


Here without loss of generality, we assume that  the rank of matrix $\Hm \Sigmam_{X\!X} \Hm^{\sf T}$ is equal to $n$, which implies that $\Sigmam_{X\!X}$ is full rank.
The rationale for this assumption comes from the observability check set by the operator, which guarantees that $\Hm$ is a full column rank matrix with $m \geq n$ for the state estimation procedure.
As a result, it holds that
$\textnormal{rank} (\Hm \Sigmam_{X\!X} \Hm^{\sf T}) = \textnormal{rank} (\Sigmam_{X\!X})$.


\subsection{Distribution of the Data Integrity}

To characterize the performance of the attacks, we obtain an equivalent expression for the performance in (\ref{Equ:obj_rv}) that shares the same distribution.

\begin{theorem} \label{Pro_EP}
The data integrity performance of the attack constructed using the sample covariance matrix is equivalent in distribution to the random variable given by
\begin{IEEEeqnarray}{ll}\label{Equ:EP}
F  \overset{d}{=}   \textnormal{\trace}\left(\Zm^{\sf T}\!\left(\tilde{\Lambdam} + \Id\right)^{-1} \tilde{\Lambdam} \Zm \right) -\log \left|\Zm^{\sf T} \tilde{\Lambdam}\Zm +\Id\right| +  \log \left| \tilde{\Lambdam} + \Id \right|, \IEEEeqnarraynumspace \middlesqueezeequ
\end{IEEEeqnarray}
where
 $\overset{d}{=}$ denotes equivalence in distribution;
$\Zm $ is an $n\times (k-1)$ matrix whose entries are zero-mean i.i.d. Gaussian random variables with variance $\frac{1}{k-1}$; 
$\tilde{\Lambdam} \eqdef \frac{1}{\sigma^2}\Lambdam \in \RR^{n\times n}$, in which $\Lambdam$ is the diagonal matrix formed with the non-zero eigenvalues of $\Hm \Sigmam_{X\!X} \Hm^{\sf T}$.
\end{theorem}

\begin{proof}
Note that
\begin{IEEEeqnarray}{rl}
  F
&  \overset{d}{=}   \trace\left(\Sigmam_{Y\!Y}^{-\!1}\Vm \Lambdam_{s}^{\frac{1}{2}} \Zm_{m} \Zm_{m}^{\sf T} \Lambdam_{s}^{\frac{1}{2}}\Vm^{\sf T}\right) + \log|\Sigmam_{Y\!Y}| \IEEEnonumber\\
& \hspace{9em} -\log \left|\Vm \Lambdam_{s}^{\frac{1}{2}} \Zm_{m} \Zm_{m}^{\sf T} \Lambdam_{s}^{\frac{1}{2}}\Vm^{\sf T}+\sigma^{2}\Id\right|   \squeezeequ \IEEEeqnarraynumspace \label{Pro_P_3}\\
&  \overset{d}{=}    \trace\left(\left(\Lambdam_{s} + \sigma^2 \Id\right)^{-1} \Lambdam_{s}  \Zm_{m} \Zm_{m}^{\sf T} \right)  +\log \left| \frac{\Lambdam_{s}}{\sigma^2} + \Id \right| \squeezeequ \IEEEnonumber  \\
& \hspace{13.5em}  -\log \left|\frac{\Lambdam_{s}}{\sigma^2} \  \Zm_{m} \Zm_{m}^{\sf T} +\Id\right|  \label{Pro_P_5} \squeezeequ \IEEEeqnarraynumspace\\
& \ \overset{d}{=}  \trace\left(\Zm^{\sf T}\!\left(\tilde{\Lambdam} + \Id\right)^{-1} \tilde{\Lambdam} \Zm \right) -\log \left|\Zm^{\sf T} \tilde{\Lambdam}\Zm +\Id\right| +  \log \left| \tilde{\Lambdam} + \Id \right|, \supersqueezeequ  \IEEEeqnarraynumspace \label{Pro_P_7}
\end{IEEEeqnarray}
where (\ref{Pro_P_3}) follows from the fact that $(k\hspace{0.06em}\!-\hspace{0.06em}\!1)\Sigmam_{\tilde{A}\!\tilde{A}} \!=\! (k\hspace{0.06em}\!-\hspace{0.06em}\!1)\Hm\Sm_{X\!X}\Hm^{\sf T} \!\sim\! \Wc_{m} (k-1, \Hm\Sigmam_{X\!X}\Hm^{\sf T})$, so it holds that
\begin{equation}
\Hm\Sm_{X\!X}\Hm^{\sf T} \overset{d}{=} \Vm \Lambdam_{s}^{\frac{1}{2}} \Zm_{m} \Zm_{m}^{\sf T} \Lambdam_{s}^{\frac{1}{2}}\Vm^{\sf T},
\end{equation}
in which $\Lambdam_{s}$ and $\Vm$ are the matrix of eigenvalues and the unitary matrix of corresponding eigenvectors, respectively, of $\Hm \Sigmam_{X\!X} \Hm^{\sf T}$, and $\Zm_{m}$ is a matrix of dimension $m \times (k-1)$ whose entries are zero-mean i.i.d. Gaussian random variables with variance $\frac{1}{k-1}$;
Given the fact that $\Sigmam_{Y\!Y} = \Hm \Sigmam_{X\!X} \Hm^{\sf T} + \sigma^2 \Id$ shares the same eigenvectors as $\Hm \Sigmam_{X\!X} \Hm^{\sf T}$ and $\log|\Sigmam_{Y\!Y}| = \log|\Lambdam_{s}+ \sigma^2\Id| $,
 (\ref{Pro_P_5}) follows from applying the cyclic permutation to the trace term in (\ref{Pro_P_3}) and applying the Sylvester's determinant identity for the logarithm of the determinant term in (\ref{Pro_P_3});
(\ref{Pro_P_7}) follows from the fact $\Lambdam_{s}$ is a rank deficient matrix with rank $n$ and applying the cyclic permutation for the trace term and the logarithm determinant term.
This completes the proof. 
\end{proof}

\subsection{Asymptotic Behaviors of Matrices}

To obtain the asymptotic performance, the asymptotic behavior of diagonal matrix $\tilde{\Lambdam} \in \RR^{n\times n}$ needs to be defined.
Given the definition of $\tilde{\Lambdam}$ in Theorem \ref{Pro_EP}, the asymptotic behavior of $\tilde{\Lambdam}$ can be obtained by defining the asymptotic behavior of $\Hm$ and $\Sigmam_{X\!X}$. 
Increasing the number of buses and transmission lines in the power system leads to the increase in the dimensions of $\Hm$ and $\Sigmam_{X\!X}$, but the values of the additive entries in $\Hm$ and $\Sigmam_{X\!X}$ are determined by the arrangement and the admittance of the transmission lines that connect the added bus with the existing buses.
To that end, there is no general model to characterize the behavior of $\Hm$ and $\Sigmam_{X\!X}$ when the dimensions increase.
That being the case,  we choose to define the asymptotic behavior of $\tilde{\Lambdam} $ directly.


Let $n_{0}$ denote the number of state variables within the practical power system that we are analysing and $\tilde{\Lambdam}_{n_0} \in \RR^{n_0\times n_0}$ denote the corresponding $\tilde{\Lambdam}$ in this system.
For example, when the voltage angles of the buses are chosen to be the state variables, there are $29$ state variables for the IEEE 30-Bus test system, 
 which implies that $n_{0} = 29$.
As a result, there are $29$ positive eigenvalues of the matrix $\Hm \Sigmam_{X\!X}\Hm^{\sf T}$ and $\tilde{\Lambdam}_{n_0}$ is of dimension $29 \times 29$.
The empirical cumulative distribution function (c.d.f.) of the diagonal elements of $\tilde{\Lambdam}_{n_0}$ is given by
\begin{IEEEeqnarray}{c}\label{Equ:Eig_Distribution}
\textnormal{\textsf{F}}_{\tilde{\mathbf{\Lambda}}_{n_0}}^{n_0}(x) = \frac{\sum_{i=1}^{n_{0}}\mathbbm{1}_{\left\{ \lambda_{i} (\tilde{\mathbf{\Lambda}}_{n_0})  \leq x)\right\}}}{n_{0}},
\end{IEEEeqnarray}
which is obtained from the parameters of the power system.

For the asymptotic scenario, we define 
\begin{IEEEeqnarray}{c}\label{Equ:Eig_Distribution_asymp}
\tilde{\mathbf{\Lambda}} = \tilde{\Lambdam}_{n_0} \otimes  \Id , 
\end{IEEEeqnarray}
where $\otimes$ is the Kronecker product.
Under this setting, the dimension of $\tilde{\mathbf{\Lambda}}$ is $n \times n$ with $n=ln_0$, in which $l$ is the dimension of the identity matrix $\Id$.
As a result, when $l\to \infty$ in (\ref{Equ:Eig_Distribution_asymp}),  the AED of $\tilde{\Lambdam}$, i.e. $\textnormal{\textsf{F}}_{\tilde{\mathbf{\Lambda}}} (x)$, is given by 
\begin{IEEEeqnarray}{c}\label{Equ:AED_tildeLambda}
\textnormal{\textsf{F}}_{\tilde{\mathbf{\Lambda}}} (x) = \lim_{n \to \infty}\frac{\sum_{i=1}^{n}\mathbbm{1}_{\left\{ \lambda_{i} (\tilde{\mathbf{\Lambda}})  \leq x\right\}}}{n} = \frac{\sum_{i=1}^{n_{0}}\mathbbm{1}_{\left\{ \lambda_{i} (\tilde{\mathbf{\Lambda}}_{n_0}) \leq x\right\}}}{n_{0}}, \squeezeequ \IEEEeqnarraynumspace
\end{IEEEeqnarray}
which states that the AED of $\tilde{\Lambdam}$ is the same as the empirical c.d.f. of the eigenvalues of $\tilde{\Lambdam}_{n_0} $.

Under the asymptotic setting in (\ref{Equ:Eig_Distribution_asymp}), we also have that
\begin{IEEEeqnarray}{c} \label{Equ:Delta_constant}
\log |\tilde{\Lambdam}+\Id| = \frac{n}{n_0} \log |\tilde{\Lambdam}_{n_0}+\Id| =n \Delta_c, 
\end{IEEEeqnarray}
where 
\begin{IEEEeqnarray}{c}\label{Equ:Delta_c}
\Delta_c = \frac{\log |\tilde{\Lambdam}_{n_0}+\Id|}{n_0}.
\end{IEEEeqnarray}
Given the fact that $\tilde{\Lambdam}_{n_0}$ and $n_0$ are determined by the power system, $\Delta_c$ is a constant for the asymptotic scenario. 


\subsection{Asymptotic Ergodic Data Integrity}

The following theorem provides the asymptotic characterization of the ergodic performance of the attacks constructed using the sample covariance matrix.

\begin{theorem}\label{Theorem:Asymp_performance}
Let $k \rightarrow\infty$ with $\frac{n}{m}\rightarrow\alpha$
 and $\frac{k-1}{n}\rightarrow\beta $, then the ergodic data integrity of the stealth attacks given by
\begin{IEEEeqnarray}{ll}\label{Equ:f_n}
&\bar{F}_n \eqdef \frac{1}{n} F \label{Equ:RMT_Performance_def}
\end{IEEEeqnarray}
converges almost surely to 
\begin{IEEEeqnarray}{c}\label{Equ:f_inf}
\bar{F}_{\infty} \eqdef \frac{1}{2} \left( 
 \lambda_{\Zm^{\sf T}\!\left(\tilde{\mathbf{\Lambda}} + \Id\right)^{-1} \tilde{\mathbf{\Lambda}}\Zm} - \log\left( 1+ \lambda_{\Zm^{\sf T} \tilde{\mathbf{\Lambda}}\Zm}\right) + \Delta_c 
\right)
 \IEEEeqnarraynumspace
\end{IEEEeqnarray}
with
\begin{IEEEeqnarray}{c}
\EE \left[\bar{F}_{\infty}\right]  = \frac{1}{2}\Big(\!\Theta+ \Delta_c \!\Big) - \frac{1}{2}\left(\Vc_{ \tilde{\mathbf{\Lambda}}}\left(\eta \right) - \beta \log \eta  + \beta \left(\eta -1 \right) \log e \right), \label{Equ:Erg_Performance} \IEEEeqnarraynumspace \middlesqueezeequ \label{Equ:f_infty}
\end{IEEEeqnarray}
where $\lambda_{\mathbf{A}}$ is the unordered eigenvalue of $\mathbf{A}$, $\Delta_c$ is defined in (\ref{Equ:Delta_c}),
\begin{equation}
\Theta\eqdef\lim_{n\to\infty}\frac{1}{n}\textnormal{\trace}\left(\left(\tilde{\Lambdam} + \Id\right)^{-1} \tilde{\Lambdam}  \right) = \EE \left[ \frac{\tilde{\Lambda}}{\tilde{\Lambda}+1}\right]
\end{equation}
with $\tilde{\Lambda}$ denoting a random variable distributed as the AED of $\tilde{\mathbf{\Lambda}}$ in (\ref{Equ:AED_tildeLambda}), 
and $\eta$ is the abbreviation for the $\eta$-transform of $\Zm^{\sf T} \tilde{\mathbf{\Lambda}}\Zm$ that is solved from (\ref{Equ:ST_HTH_1}) for $\gamma = 1$.
\end{theorem}

\begin{proof}
Starting from (\ref{Equ:EP}) and (\ref{Equ:RMT_Performance_def}), we have that
\begin{IEEEeqnarray}{ll}
 \bar{F}_{\infty}  
&  = \lim_{n \to \infty}\frac{1}{n}F \\
 &   = \frac{1}{2n}\left(\!\trace\left(\!\Zm^{\sf T}\!\left(\!\tilde{\Lambdam} + \Id\!\right)^{\!-\!1} \tilde{\Lambdam} \Zm \!\right) \!-\log \left|\Zm^{\sf T} \tilde{\Lambdam}\Zm +\Id\right|  \!+  \log \left| \tilde{\Lambdam} + \Id \right|  \!\right)  \label{Equ:WPerf_1}\IEEEeqnarraynumspace  \Tsupersqueezeequ \\
 & \to \frac{1}{2} \left( \lambda_{\Zm^{\sf T}\left(\tilde{\mathbf{\Lambda}} + \Id\right)^{-1} \tilde{\mathbf{\Lambda}}\Zm} - \log\left( 1+ \lambda_{\Zm^{\sf T} \tilde{\mathbf{\Lambda}}\Zm}\right) + \Delta_c \right). \label{Equ:WPerf_2}  \IEEEeqnarraynumspace 
\end{IEEEeqnarray}

We now characterize the ergodic performance, i.e. the expected value of (\ref{Equ:WPerf_2}).
Following the same procedure as in (\ref{Equ:var_trace}), we have 
\begin{IEEEeqnarray}{rl} \label{Equ:ergodic_1}
\EE \hspace{-0.25em}\left[ \lambda_{\Zm^{\sf T}\left(\tilde{\mathbf{\Lambda}} + \Id\right)^{-1} \tilde{\mathbf{\Lambda}}\Zm}\right]  & = \! \lim_{n \to \infty} \hspace{0.1em} \frac{1}{n} \hspace{0.1em}\EE \left[\!\trace \left(\left(\tilde{\mathbf{\Lambda}} + \Id\right)^{-1} \tilde{\mathbf{\Lambda}}\Zm \Zm^{\sf T}\right) \!\right] 
 = \hspace{0.05em}  \Theta. \IEEEeqnarraynumspace \Tsupersqueezeequ
\end{IEEEeqnarray}
Note that $\EE \left[ \log(1+ \lambda_{\Zm^{\sf T} \tilde{\mathbf{\Lambda}}\Zm}\right])$ is the Shannon transform of the AED of $\Zm^{\sf T} \tilde{\mathbf{\Lambda}}\Zm$, which is characterized in Theorem \ref{Theorem:ST}.
The proof is completed by taking (\ref{Equ:ST_1}) and (\ref{Equ:ergodic_1}) into (\ref{Equ:WPerf_2}). 
\end{proof}


It follows from Theorem \ref{Theorem:Asymp_performance} that we need to obtain the $\eta$-transform of $\Zm^{\sf T} \tilde{\mathbf{\Lambda}}\Zm$ from (\ref{Equ:ST_HTH_1}) to finalize the asymptotic characterization of the ergodic performance.
The following proposition shows that (\ref{Equ:ST_HTH_1}) has a unique solution of $\eta$.

\begin{proposition}\label{Pro_Solvable}
As a function of $\eta$,  (\ref{Equ:ST_HTH_1}) in Theorem \ref{Theorem:ST} has a unique solution.
\end{proposition}
\begin{proof}
Note that in (\ref{Equ:ST_HTH_1}) we have
\begin{IEEEeqnarray}{l}
\eta_{\mathbf{T}}(\gamma \eta) = \EE \left[ \frac{1}{1+\gamma\eta T}\right],
\end{IEEEeqnarray}
where $\eta_{\mathbf{L}^{\sf T} \Tm \mathbf{L}}(\gamma)$ is abbreviated as $\eta$ and the expectation is over random variable $T$ whose distribution is the AED of $\Tm$.
After some algebraic manipulation, (\ref{Equ:ST_HTH_1})  can be expressed as
\begin{align}
\beta \eta - \EE \left[ \frac{1}{1+\gamma\eta T}\right] = \beta - 1. \label{Equ:eta_Wishart_1}
\end{align}
Note that $\gamma >0$, $T \in \RR^{+}$, and the range of $\eta$ is within the interval $(0,1]$, see Definition \ref{Def:eat}, the left-hand term of (\ref{Equ:eta_Wishart_1}) is a monotonically increasing function of $\eta \in (0,1]$ and its range contains the value $ \beta - 1$. This completes the proof.

\end{proof}


\section{Variance Bounds of the Asymptotic Data Integrity} \label{Sec:variance}
In the previous section, Theorem \ref{Theorem:Asymp_performance} characterizes the ergodic performance of the attack, which is described via the equivalent distribution obtained in Theorem \ref{Pro_EP}.
However, the ergodic performance defined there only yields the average performance of the attacks. 
The variance of the performance provides insight into the probability that the performance concentrates around the averaged performance.
In the following, we propose the lower and upper bounds for the variance of the asymptotic performance of attack. 

Note that it would be ideal to characterize the distribution of $F$ in (\ref{Equ:obj_rv}) in a closed-form manner via Theorem \ref{Theorem_Variance}. 
However,  the diagonal matrix within the trace term in (\ref{Equ:EP}), i.e. ${(\tilde{\Lambdam} + \Id)^{-1} \tilde{\Lambdam}}$, is different from the one within the logarithm of the determinant term, i.e. from $\tilde{\Lambdam}$.
So here we choose to bound the variance of the performance. 

Given the fact that $\Delta_c$ defined in (\ref{Equ:Delta_constant}) and (\ref{Equ:Delta_c}) is a constant term, we only need to characterize the variance introduced by the first two terms on the left-hand side of (\ref{Equ:EP}). 
Using the equivalent distribution in Theorem \ref{Pro_EP} and further denoting 
\begin{IEEEeqnarray}{rl}
F_a &\eqdef \ \trace\left(\Zm^{\sf T}\!\left(\tilde{\Lambdam} + \Id\right)^{-1} \tilde{\Lambdam} \Zm \right) \label{Equ:f_1^1}\\
F_b &\eqdef  \ \log \left|\Zm^{\sf T} \tilde{\Lambdam}\Zm +\Id\right| \label{Equ:f_1^2},\IEEEeqnarraynumspace \squeezeequ
\end{IEEEeqnarray}
 the variance of $F$ in (\ref{Equ:obj_rv}) is given by 
\begin{IEEEeqnarray}{c} \label{Equ:bound_1}
 \textnormal{var} \left[ F\right] =  \textnormal{var} \left[ F_a\right] + \textnormal{var} \left[ F_b\right] - 2\rho \left( F_a, F_b\right)\sqrt{\textnormal{var} \left[F_a \right]}\sqrt{\textnormal{var} \left[F_b \right]} , \IEEEeqnarraynumspace \supersqueezeequ
\end{IEEEeqnarray}
where $\rho \left( \cdot, \cdot \right)$ denotes the Pearson correlation. 

We proceed by proving that $\rho \left( F_a, F_b\right) \in [0,1]$. 
\begin{lemma}\label{lemma:Postive_dep}
Let $\Zm $ is an $n\times (k-1)$ matrix whose entries are zero-mean i.i.d. Gaussian random variables with variance $\frac{1}{k-1}$. 
Let $\tilde{\Lambdam}$ be an $n \times n$ diagonal and nonnegative random matrix, which is independent of $\Zm$. 
Then it holds that 
\begin{IEEEeqnarray}{c}
0 \leq \rho \left( F_a, F_b\right) \leq 1, 
\end{IEEEeqnarray}
where $F_a$ and $F_b$ are defined in (\ref{Equ:f_1^1}) and (\ref{Equ:f_1^2}), respectively. 
\end{lemma}

\begin{proof}
Note that 
\begin{IEEEeqnarray}{c}
F_a 
\!=  \trace\left(\!\left(\tilde{\Lambdam} + \Id\right)^{-1} \tilde{\Lambdam}^{\frac{1}{2}}\hspace{0.1em}  \Zm \left( \tilde{\Lambdam}^{\frac{1}{2}}\hspace{0.1em}  \Zm\right)^{\! \sf T} \right) 
\!=\!  \sum_{i=1}^n \hspace{0.1em}  \frac{1}{1+ \lambda_{\bm{\sigma}(i)}(\tilde{\Lambdam}+\Id)} \bar{\lambda}_i , \IEEEeqnarraynumspace \supersqueezeequ \label{Equ:permutation_trace_2}
\end{IEEEeqnarray}
where 
the first equality follows from the fact that $\tilde{\Lambdam}$ is a diagonal matrix and from applying cyclical permutation to the trace term; 
 the second equality follows from \cite[6.57]{seber_matrix_2008}, in which $\sigmav \in \mathbb{N}^{n}$ is a permutation of $[1,2, \dots,n]^{\sf T}$,  and $\bar{\lambda}_i$ is the $i$-th eigenvalue of $\tilde{\Lambdam}^{\frac{1}{2}}  \Zm ( \tilde{\Lambdam}^{\frac{1}{2}}  \Zm)^{\sf T}  $. 

Furthermore, we have
\begin{IEEEeqnarray}{c}
F_b 
=  \log \left|\tilde{\Lambdam}^{\frac{1}{2}}  \Zm ( \tilde{\Lambdam}^{\frac{1}{2}}  \Zm )^{\sf T}  +\Id\right| 
=    \sum_{i=1}^n \log(1+ \bar{\lambda}_i),  \label{Equ:logdet_3}
\end{IEEEeqnarray}
where (\ref{Equ:logdet_3}) follows from Sylvester's determinant identity.

From (\ref{Equ:permutation_trace_2}) and (\ref{Equ:logdet_3}), it is easy to show that both $F_a$ and $F_b$ are coordinatewise monotonically increasing functions of the vector $\bar{\Lambdam} = \left[ \bar{\lambda}_1 , \dots, \bar{\lambda}_n\right]^{\sf T}$.
Note that $\bar{\Lambdam}$ is the vector of eigenvalues of a Wishart random matrix distributed as $\Wc_{n}(k-1,\tilde{\Lambdam})$. 
It is proved in \cite[Theorem 3]{tsai_2019_ordering} that the distribution of the eigenvalues of a Wishart random matrix is multivariate totally positive of order 2 ($\textnormal{MTP}_{\textnormal{2}}$).
Adding the fact that two coordinatewise monotonically increasing or decreasing functions of a vector whose distribution is $\textnormal{MTP}_{\textnormal{2}}$ are positively correlated \cite[(1.9)]{karlin_1980_classes}, the conclusion that $0 \leq \rho \left( F_a, F_b\right) \leq 1$ holds for any realizations of $\tilde{\Lambdam}$. 
The theorem follows from the independence between $\tilde{\Lambdam}$ and $\Zm$.
 \end{proof}
 
 
Using the result in Lemma \ref{lemma:Postive_dep}, the upper bound in (\ref{Equ:bound_1}) is transformed into 
 \begin{IEEEeqnarray}{c} \label{Equ:bound_2}
 \left( \sqrt{\textnormal{var} \left[ F_a\right]} - \sqrt{\textnormal{var} \left[F^b \right]} \right)^2 \leq
 \textnormal{var} \left[ F\right] \leq
\textnormal{var} \left[ F_a \right]+ \textnormal{var} \left[ F_b  \right]
, \IEEEeqnarraynumspace \squeezeequ
\end{IEEEeqnarray}
where the upper bound is achieved when $\rho \left( F_a, F_b\right) = 0$, and the lower bound is achieved when $\rho \left( F_a, F_b\right) = 1$.

 The following theorem provides a lower bound and an upper bound for the variance of the performance. 

 \begin{theorem} \label{Theorem:Var_bound}
The variance of the data integrity of the attacks constructed using sample covariance matrix, i.e. $\textnormal{var} \left[ F \right]$ with $F$ defined in (\ref{Equ:obj_rv}), is bound by 
\begin{IEEEeqnarray}{c}
\frac{1}{4}\Big( \sqrt{\hspace{-0.1em}\textnormal{var} [ F_a]} -\hspace{-0.1em} \sqrt{\textnormal{var} [F_b ]} \Big)^2 
\leq \textnormal{var} \left[ F\right] \leq
 \hspace{0.15em}\frac{1}{4} \hspace{0.15em} \left(\textnormal{var} \left[ F_a \right]+ \textnormal{var} \left[ F_b  \right]\right) \IEEEeqnarraynumspace \supersqueezeequ
\end{IEEEeqnarray}
with
\begin{IEEEeqnarray}{rl}
\textnormal{var} \left[ F_a \right] &= \frac{2}{\beta} \EE\left[ \left(\frac{\tilde{\Lambda}}{\tilde{\Lambda}+1}\right)^2\right] \label{Equ:var_1}\\
\textnormal{var} \left[ F_b  \right] & = -\log \left( 1 - \frac{1}{\beta} \EE \left[ \left( \frac{ \tilde{\Lambda}\eta_{\mathbf{Z}^{\sf T}  \tilde{\mathbf{\Lambda} } \mathbf{Z}}(1)}{1+\tilde{\Lambda}\eta_{\mathbf{Z}^{\sf T} \tilde{\mathbf{\Lambda} }\mathbf{Z}}(1)}\right)\right]\right), \label{Equ:var_2}
\end{IEEEeqnarray}
where the expectation is over the nonnegative random variable $\tilde{\Lambda}$, whose distribution is the AED of $\tilde{\Lambdam}$ defined in (\ref{Equ:AED_tildeLambda}); 
 and the value of $\eta_{\mathbf{Z}^{\sf T}  \tilde{\mathbf{\Lambda} } \mathbf{Z}}(1) $ is solved from (\ref{Equ:ST_HTH_1}), which is proved to be always solvable in Proposition \ref{Pro_Solvable}.
\end{theorem}

\begin{proof}
Note that $F_a - F_b$ is the only term that introduces the randomness into the objective, as described in (\ref{Equ:bound_1}).
As a result, the theorem follows directly from combining the results in Theorem \ref{Theorem:ST_Variance} and Theorem \ref{Theorem:var_trace} with  (\ref{Equ:bound_2}).
\end{proof}

For the variance bounds in Theorem \ref{Theorem:Var_bound}, the difference between the upper bound and lower bound is further upper bounded by $\frac{1}{2}\sqrt{\frac{2}{\beta}}\sqrt{-\log(1-\frac{1}{\beta})}$, regardless of the distribution of $\tilde{\Lambda}$.
In particular, the difference between the bounds is smaller than $0.0726$ when $\beta = 10$ and is smaller than $0.0071$ when $\beta = 100$ for the natural logarithm case.

\section{Numerical Simulation} \label{Sec:Numerical_Simulation}
In this section, we present simulations to evaluate the performance of the attacks constructed using the sample covariance matrix in practical state estimation settings. 
In particular, we use the IEEE 30-Bus and 118-Bus test systems, whose parameters and topology are obtained from MATPOWER \cite{zimmerman_matpower:_2011}.
We assume a DC state estimation scenario \cite{abur_power_2004, grainger_power_1994}, for which the bus voltage angle is chosen to be the state variables.


It is worth mentioning that our results in this paper hold for any covariance matrix of the state variable and for any $\beta >0$, regardless of the structure of the matrix.
In the simulations, the covariance matrix of the state variables is assumed to be a Toeplitz matrix with exponential decay parameter $r \in [0,1]$, i.e. 
$\Sigmam_{X\!X}=[s_{ij}=r^{|i-j|}; i, j =1, 2, \ldots, n]$, 
where the exponential decay parameter $r$ determines the correlation strength between different entries of the state variable vector.
And for $\beta$, we assume that the number of samples available to the attacker is larger than the dimension of the state variables, i.e. $k-1 \geq n$ or $\beta \geq 1$. 
This guarantees that the objective function in (\ref{Equ:obj_rv_0}) is always computable for the nonasymptotic case, i.e. for the practical IEEE test systems.
The Signal-to-Noise Ratio (SNR) of the power system is defined as
\begin{equation}
\textnormal{SNR} \eqdef 10\log_{10}\left(\frac{\trace{(\Hm\Sigmam_{X\!X}\Hm^{\sf T}})}{m\sigma^2}\right). 
\end{equation}

\begin{figure}[t!]
\centering
\includegraphics[scale=0.4]{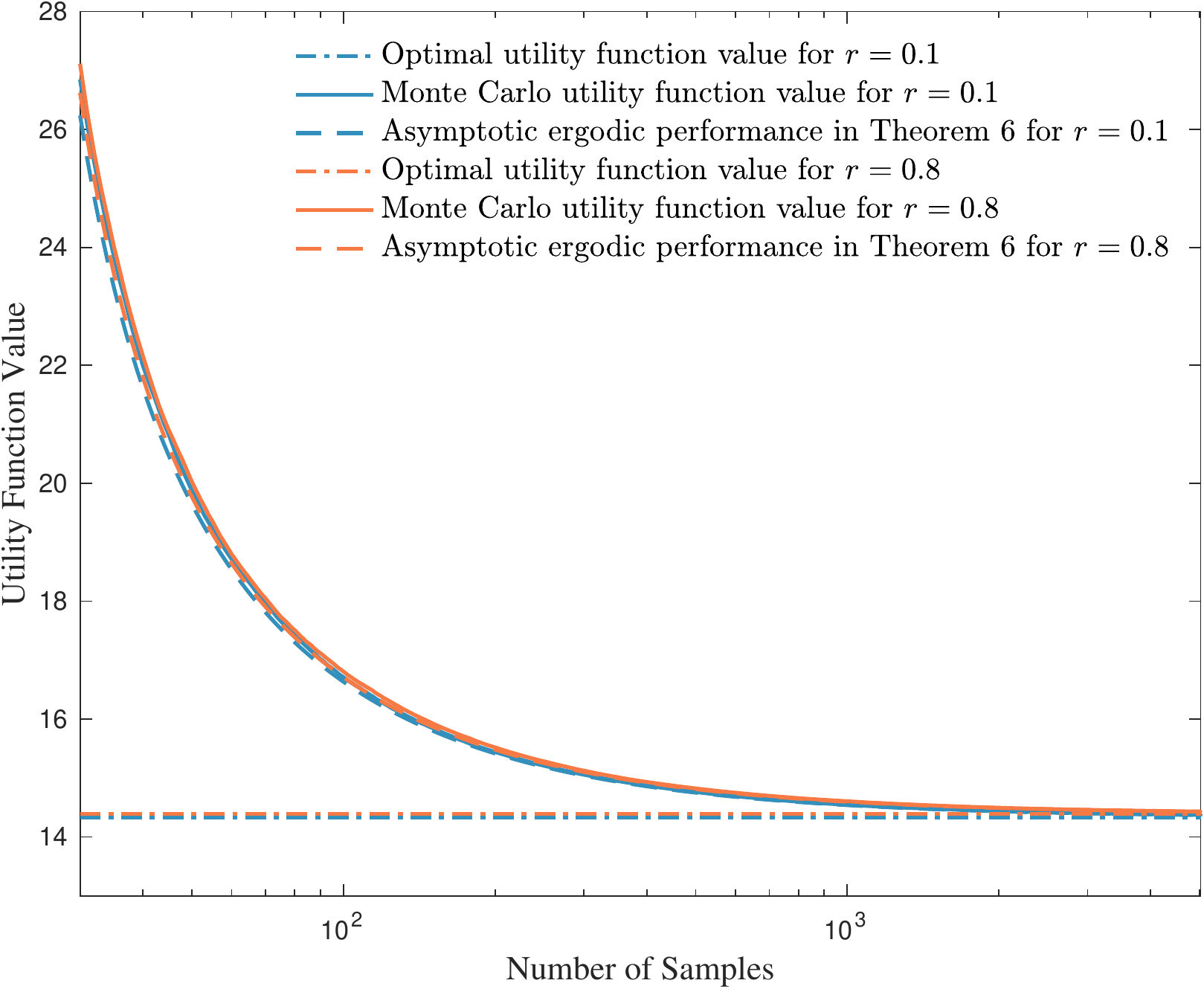}
\caption{ Performance of the  asymptotic ergodic data integrity in Theorem \ref{Theorem:Asymp_performance} for $r = 0.1$ and $r = 0.8$ when $\textnormal{SNR} = 30 \hspace{0.1em}\textnormal{dB}$ on IEEE 30-Bus test system.}
\label{Fig:ST_Rho_30_SNR30}
\end{figure}

\begin{figure}[t!]
\centering
\includegraphics[scale=0.4]{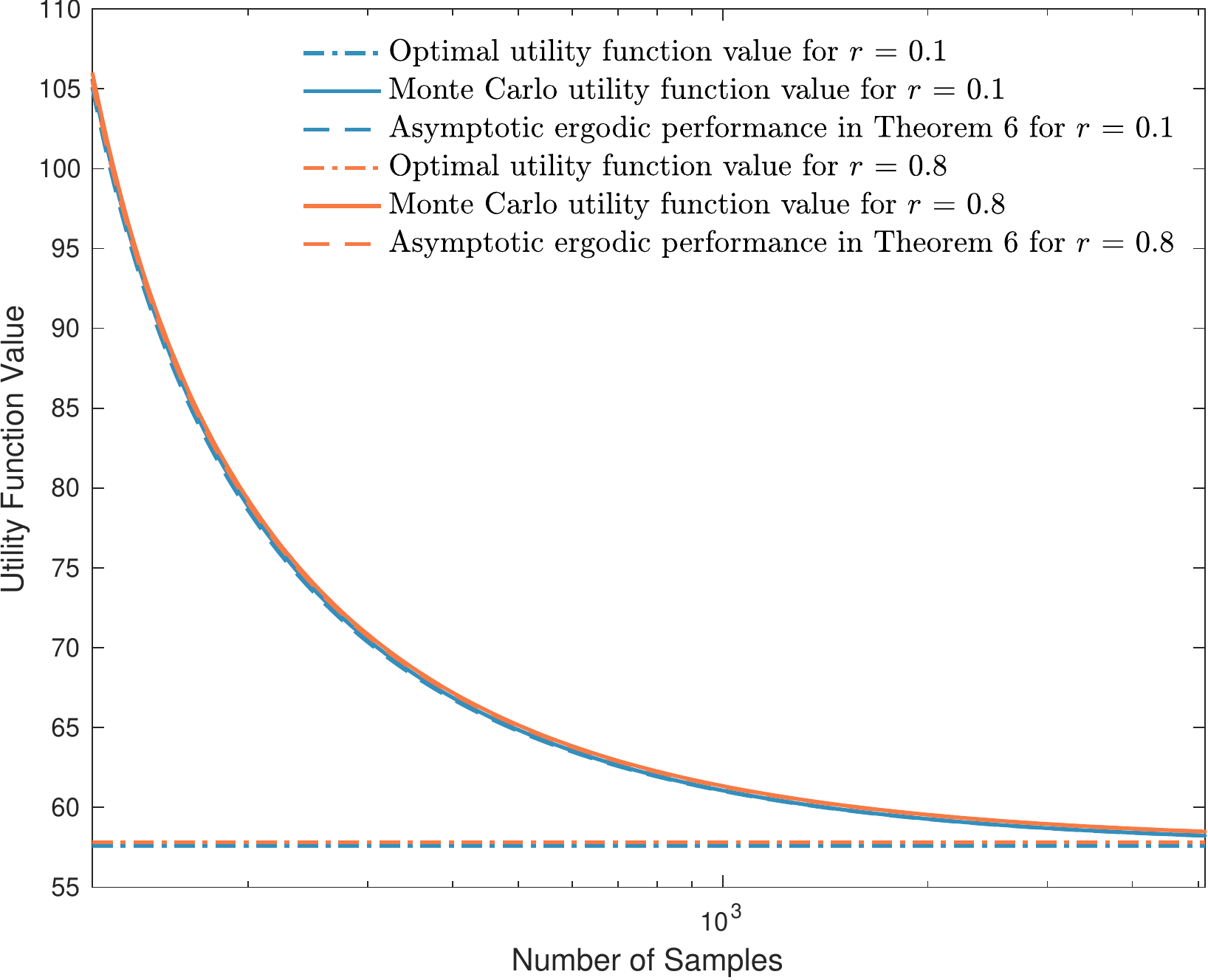}
\caption{ Performance of the  asymptotic ergodic data integrity in Theorem \ref{Theorem:Asymp_performance} for $r = 0.1$ and $r = 0.8$ when $\textnormal{SNR} = 30 \hspace{0.1em}\textnormal{dB}$ on IEEE 118-Bus test system.}
\label{Fig:ST_Rho_118_SNR30}
\end{figure}

\begin{figure}[t!]
\centering
\includegraphics[scale=0.4]{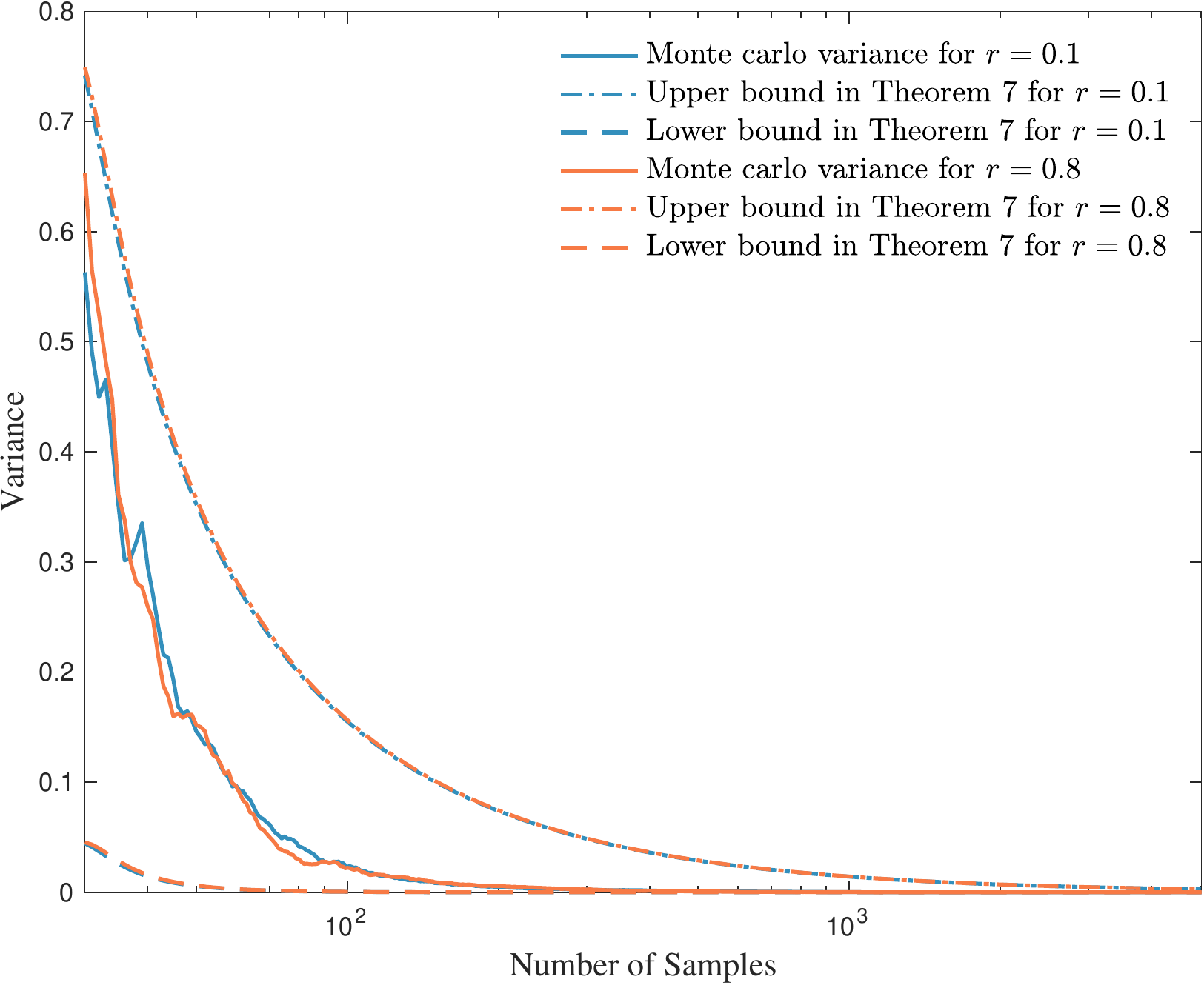}
\caption{ Performance of the asymptotic ergodic data integrity in Theorem \ref{Theorem:Var_bound} for $r = 0.1$ and $r = 0.8$ when $\textnormal{SNR} = 30 \hspace{0.1em}\textnormal{dB}$ on IEEE 30-Bus test system.}
\label{Fig:Bound_Var_30_SNR30}
\end{figure}

\begin{figure}[t!]
\centering
\includegraphics[scale=0.4]{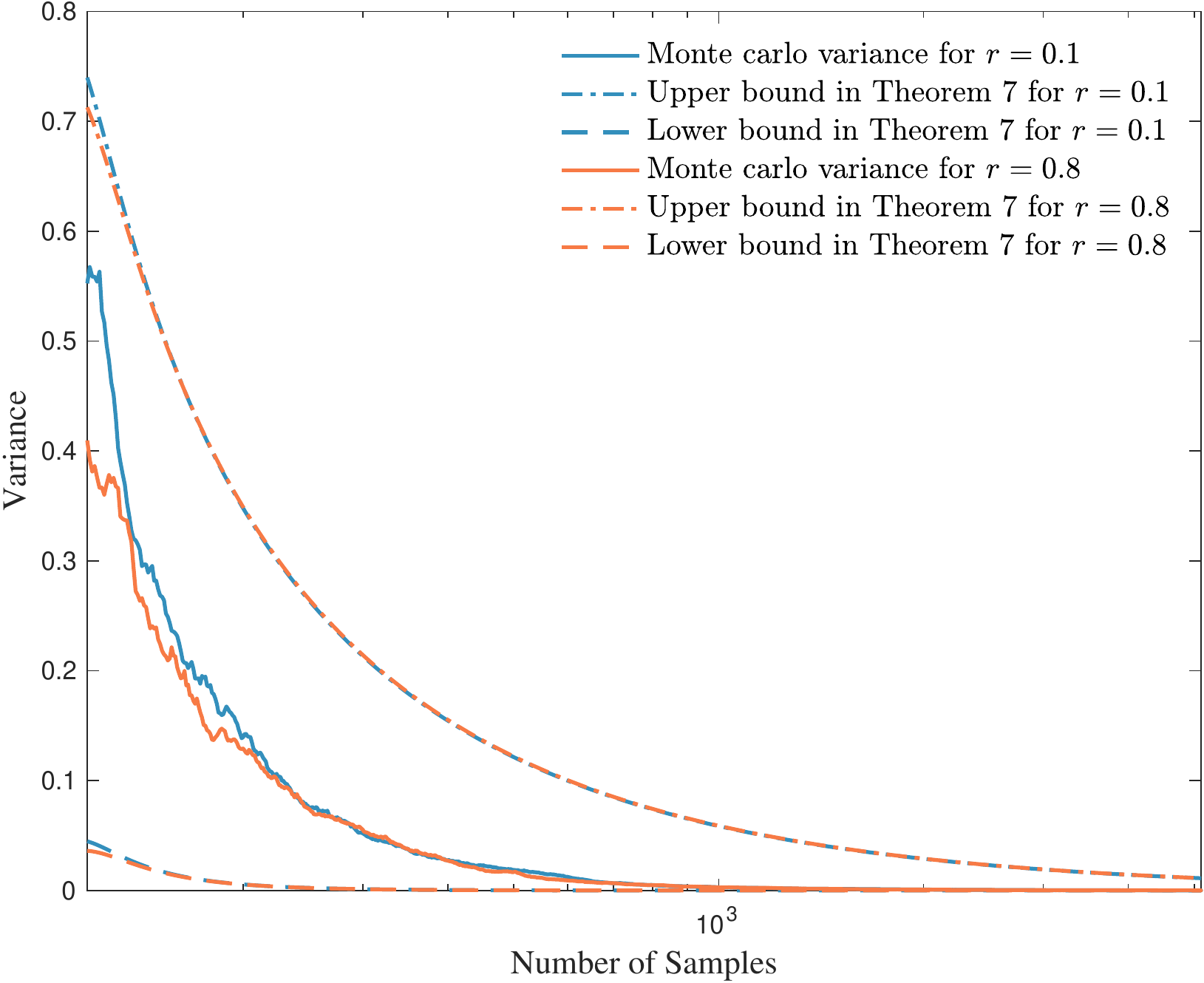}
\caption{ Performance of the asymptotic ergodic data integrity in Theorem \ref{Theorem:Var_bound} for $r = 0.1$ and $r = 0.8$ when $\textnormal{SNR} = 30 \hspace{0.1em}\textnormal{dB}$ on IEEE 118-Bus test system.}
\label{Fig:Bound_Var_118_SNR30}
\end{figure}

Fig. \ref{Fig:ST_Rho_30_SNR30} depicts the performance of the asymptotic ergodic data integrity in Theorem \ref{Theorem:Asymp_performance} on IEEE 30-Bus test system for $r = 0.1$ and $r = 0.8$ when $\textnormal{SNR} = 30 \hspace{0.1em}\textnormal{dB}$,  in which the Monte Carlo performance value is the averaged performance through one thousand realizations, and the optimal utility function value is the utility function value when the attacker has perfect knowledge about the system.
It is found that when the number of samples increases, the performance of the attacks constructed with the sample covariance matrix converges to the optimal value. 
More importantly, the asymptotic characterization approximates the non-asymptotic case described by the real power system well. 
It is worth to mentioning that the superb approximation still holds when the SNR changes.
The same phenomenon is observed for the simulation on IEEE 118-Bus test system, which is shown in Fig. \ref{Fig:ST_Rho_118_SNR30}.

Fig. \ref{Fig:Bound_Var_30_SNR30} depicts the performance of the bounds that are proposed in Theorem \ref{Theorem:Var_bound} for the IEEE 30-Bus test system when $r = 0.1$ and $r = 0.8$ with $\textnormal{SNR} = 30 \hspace{0.1em}\textnormal{dB}$. 
Compared with the case when $\textnormal{SNR} = 10 \hspace{0.1em}\textnormal{dB}$, it is found that the bounds, especially the lower bound, is tighter when the $\textnormal{SNR}$ value is high.
Furthermore, the variance obtained by the Monte Carlo approach is closer to the upper bound when the number of samples is small compared with the dimension of the system, i.e. $\beta$ is small, and when $\textnormal{SNR}$ is high.
The same phenomenon is observed for the simulation on IEEE 118-Bus test system, which is shown in Fig. \ref{Fig:Bound_Var_118_SNR30}.
Interestingly, the variance and the upper and lower bounds are comparable for both the IEEE 30-Bus test system and the IEEE 118-Bus test system. 
This suggests that the tightness of the bounds does not change with the size of the system.

It is worth mentioning that the conclusions in the preceding context also hold for the linearized AC model in (\ref{Equ:Hm}).
From a practical point of view, the results indicate that when the number of samples in the training set is at least 10 times larger than the dimension of the vector of state variables, the performance of the stealth attacks is close to that of the attack construction with perfect knowledge.
 Moreover, for that case the variance is smaller than $0.1$, which suggests that the attack performance is close to the optimal case for most training data set realizations. 
 This insight provides a guideline for operators on how much historical data is safe to share between different stakeholders in the power system.
For instance, the historical data, such as voltage angle and magnitude, owned by Transmission System Operators or by Distribution System Operators might pose a risk depending on the size of the data set determined by $\beta$. Our analytical results provide a quantitative framework to assess the risk of sharing historical data in platforms such as the data exchange hubs of the National Regulatory Authority  \cite{jenssen_2017_data}.

\section{Conclusion}\label{Sec:Conclusion}

In this paper, the learning requirements for information-theoretic DIAs have been analyzed using asymptotic RMT tools. 
Specifically, in this framework the attacker learns the second-order statistics of the state variables from a limited number of past realizations of the state variables and constructs the attacks using the estimated statistics. 
Since the sample covariance matrix is a random matrix, the performance of the attacks using the estimated statistics is a random variable. 
The ergodic performance of the attacks using the estimated statistics has been characterized in closed-form and the variance of the performance is bounded to obtain insight into the distribution of the performance.
It is observed from the numerical simulations that the non-asymptotic ergodic performance exhibits an exponential convergence to the asymptotic case, and therefore, the asymptotic characterization provides practical insight into the performance of stealth attacks even with small datasets.

%
%
%
%
%
%
%



\bibliographystyle{IEEEbib}
\bibliography{reference}

\end{document}